\documentclass{article}
\usepackage{amssymb}
\usepackage{makeidx}
\usepackage{amsfonts}
\usepackage{dsfont}

\begin{document}

\title{A hidden symmetry of complex spacetime and the emergence of the
standard model algebraic structure}
\author{R. Vilela Mendes\thanks{%
rvilela.mendes@gmail.com; rvmendes@fc.ul.pt;
https://label2.tecnico.ulisboa.pt/vilela/} \\
CMAFcIO, Faculdade de Ci\^{e}ncias, Univ. Lisboa}
\date{ }
\maketitle

\begin{abstract}
When spacetime is considered as a subspace of a wider complex spacetime
manifold, there is a mismatch of the elementary linear representations of
their symmetry groups, the real and complex Poincar\'{e} groups. In
particular, no spinors are allowed for the complex case. When a spin$^{h}$
structure is implemented on principal bundles in complex spacetime, one is
naturally led to an algebraic structure similar to the one of the standard
model.
\end{abstract}

\section{Introduction: Real fibers in complex spacetime}

In the past, several authors have studied extensions of physical theories to
the complex domain. Einstein in 1945 \cite{Einstein} generalized his
gravitational theory to a metric with complex components, others constructed
solutions of the complex Einstein equation \cite{Kunstatter} \cite{Plebanski}
or a complex manifold description of massless particles \cite{Penrose-2} 
\cite{Esposito-1}, etc.

A more radical hypothesis would be to assume that the spacetime manifold
itself is parametrized by complex coordinates. Explorations in this
direction have been carried out, for example \cite{Brown} \cite{Newman}, and
a more complete set of references and perspectives may be obtained from Ref.%
\cite{Esposito-2}. A important question that was left open is, if the
spacetime dimensions are indeed indexed by complex numbers, why we do not
feel it in our everyday experience. This question has been revisited for
spacetimes where coordinates take values in normed division algebras,
complex, quaternion or octonion \cite{IJMPA}. For the complex field case, a
complex manifold $\mathcal{M}$ is considered, with (local) metric $G=\left(
1,-1,-1,-1\right) $ and symmetry group $L_{\mathbb{C}}\mathcal{=}U\left(
1,3\right) $%
\begin{equation}
\Lambda ^{\dag }G\Lambda =G  \label{1.1}
\end{equation}%
$\left( \Lambda \in L_{\mathbb{C}}\right) $, the real spacetime being
considered to be a $4-$dimensional subspace $\mathcal{S}$ with Lorentzian
metric, that is, a Lorentzian fiber in the Grassmannian $Gr_{4}\left( 
\mathcal{M}\right) $ of four-dimensional frames of $\mathcal{M}$. The
symmetry group of $\mathcal{S}$ is the real Lorentz group $L_{\mathbb{R}}%
\mathcal{=}SO\left( 1,3\right) $\textit{.} Together with the complex
inhomogeneous translations, the (local) symmetry group of $\mathcal{M}$ is  
\begin{equation}
P_{\mathbb{C}}=T_{4}\circledS U\left( 1,3\right)   \label{1.2}
\end{equation}%
$P_{\mathbb{C}}$ is the group that, in the past, has been called the Poincar%
\'{e} group with real metric \cite{Barut} \cite{Ungar} \cite{Kursunoglu}. It
is a $24-$parameter complex group distinct from the $16-$parameter group
used in the analytic continuations of the S-matrix.

The nature of the matter states, that is implied by this group structure,
was obtained \cite{IJMPA} by studying the representations of the semidirect
group (\ref{1.2}). The conclusions were:

- Half-integer spin states are elementary states only for the real
Lorentzian fibers, not for the complex manifold $\mathcal{M}$, nor for the
higher normed algebras. Their absence as elementary states in the complex
manifold $\mathcal{M}$ may be physically interpreted as meaning that such
states cannot be "rotated" away from each real Lorentzian fiber to the other
fibers of the Grassmannian $Gr_{4}\left( \mathcal{M}\right) $.

- Integer-spin states are elementary states of the complex group. However a
superselection rule emerges when discrete symmetries are included \cite{dark}%
.

Whatever the point of view concerning the extension to complex spacetime, a
relevant question is: When real spacetime is embedded into a higher
dimensional division algebra structure, what properties of the higher
structure are inherited or non-inherited by the real spacetime? The absence
of spinor states for the complex $P_{\mathbb{C}}$ group is particularly
interesting. It might be that $P_{\mathbb{C}}$ is actually also the symmetry
group of the real Lorentzian fiber, but that the symmetry is implemented in
a subtler way. This second point of view leads to an analysis of when spin
structures are or are not realized in coset manifolds. And to what
additional quantum numbers appear when one insists on their implementation.

\section{Consistency of the symmetry representations and emergence of an
algebraic structure}

The generators of the Lie algebra of the complex Poincar\'{e} group $P_{%
\mathbb{C}}$ are $\left\{ M_{\mu \nu },N_{\mu \nu },K_{\mu },H_{\mu
}\right\} $ , $M_{\mu \nu }$ and $N_{\mu \nu }$ being the generators of the $%
U(1,3)$ algebra, $P_{\mu }$ and $H_{\mu }$ the generators of real and
complex translations, with commutation relations%
\begin{eqnarray}
\left[ M_{\mu \nu },M_{\rho \sigma }\right]  &=&-M_{\mu \sigma }g_{\nu \rho
}-M_{\nu \rho }g_{\mu \sigma }+M_{\nu \sigma }g_{\rho \mu }+M_{\mu \rho
}g_{\nu \sigma }  \nonumber \\
\left[ M_{\mu \nu },N_{\rho \sigma }\right]  &=&-N_{\mu \sigma }g_{\nu \rho
}+N_{\nu \rho }g_{\mu \sigma }+N_{\nu \sigma }g_{\rho \mu }-N_{\mu \rho
}g_{\nu \sigma }  \nonumber \\
\left[ N_{\mu \nu },N_{\rho \sigma }\right]  &=&M_{\mu \sigma }g_{\nu \rho
}+M_{\nu \rho }g_{\mu \sigma }+M_{\nu \sigma }g_{\rho \mu }+M_{\mu \rho
}g_{\nu \sigma }  \nonumber \\
\left[ M_{\mu \nu },P_{\rho }\right]  &=&-g_{\nu \rho }P_{\mu }+g_{\mu \rho
}P_{\nu }  \nonumber \\
\left[ M_{\mu \nu },H_{\rho }\right]  &=&-g_{\nu \rho }H_{\mu }+g_{\mu \rho
}H_{\nu }  \nonumber \\
\left[ N_{\mu \nu },P_{\rho }\right]  &=&-g_{\nu \rho }H_{\mu }-g_{\mu \rho
}H_{\nu }  \nonumber \\
\left[ N_{\mu \nu },H_{\rho }\right]  &=&g_{\nu \rho }P_{\mu }+g_{\mu \rho
}P_{\nu }  \nonumber \\
\left[ P_{\mu },H_{\rho }\right]  &=&0  \label{2.1}
\end{eqnarray}

Several authors in the past, looking for spectrum generating algebras, have
dismissed the $U(1,3)$ algebra, as useless, for not having irreducible
spinor representations. However, the fact that $O(1,3)$ has spinor
representations and $U(1,3)$ does not, is the most interesting feature of
the latter.

If however $U(1,3)$ is an invariance of Nature, in a complex spacetime
context, the apparent spectrum mismatch of the symmetries in the full space
with those of the real Lorentzian fibers has interesting consequences. The
fact that there are no linear irreducible spinor representations, does not
mean that such structures might not be present in a nonlinear realization.

Let us analyze the mismatch for the massive representations of the complex
Poincar\'{e} group $P_{\mathbb{C}}$, that is, for non-zero eigenvectors of
the $P^{2}=P_{\mu }P^{\mu }+H_{\mu }H^{\mu }$ operator. The isotropy group 
\cite{IJMPA} is $U(3)$ which one splits as $U(1)\oplus SU(3)$. Denote the
generators of $U(3)$ as $\left\{ \mathcal{G}_{i};i=1,\cdots ,9\right\}
=\left\{ R_{1},R_{2},R_{3},U_{1},U_{2},U_{3},D_{1},D_{2},D_{3}\right\} $%
\footnote{%
A representation in terms of the usual $SU\left( 3\right) $ $\lambda $%
-matrices is: $R_{1}\rightarrow i\lambda _{7};\;R_{2}\rightarrow -i\lambda
_{5};\;R_{3}\rightarrow i\lambda _{2};U_{1}\rightarrow i\lambda
_{6};U_{2}\rightarrow i\lambda _{4};U_{3}\rightarrow i\lambda
_{1};D_{1}\rightarrow i\lambda _{3};D_{2}\rightarrow \frac{i}{2}\lambda
_{8};D_{3}\rightarrow i\sqrt{\frac{2}{3}}\mathds{1}$}%
\begin{eqnarray}
R_{i} &=&\frac{1}{2}\epsilon _{ijk}M_{jk};U_{i}=\frac{1}{2}\epsilon
_{ijk}N_{jk};D_{1}=\frac{1}{2}\left( N_{11}-N_{22}\right) ;  \nonumber \\
D_{2} &=&\frac{1}{2\sqrt{3}}\left( N_{11}+N_{22}-2N_{33}\right) ;D_{3}=\frac{%
1}{\sqrt{6}}\left( N_{11}+N_{22}+N_{33}\right)   \label{2.2}
\end{eqnarray}%
the first eight being the generators of $SU\left( 3\right) $ and $\mathcal{G}%
_{9}$ the generator of $U\left( 1\right) $. In a standard $3\times 3$
representation they are trace orthogonal%
\begin{equation}
Tr\left( \mathcal{G}_{i}\mathcal{G}_{j}\right) =-2\delta _{ij}  \label{2.3}
\end{equation}

With a coset decomposition, $SU(3)$ becomes a $SO(3)$ principal bundle $%
\mathcal{C}\left( \mathcal{C}_{b},\mathcal{C}_{f},SO(3)\right) $, with base
the Riemmanian manifold $\mathcal{C}_{b}=SU(3)/SO(3)$ and fibers $\mathcal{C}%
_{f}\simeq SO(3)$. The metric in $\mathcal{C}_{b}$ is obtained from (\ref%
{2.3}). A coordinate system in $\mathcal{C}$ is obtained by choosing a point
in each coset%
\begin{equation}
\left( \omega ^{i}\right) =\exp \left( \omega ^{i}\mathcal{G}_{i};i=4,\cdots
,8\right)   \label{2.4}
\end{equation}%
and when an element $g\in SU(3)$ acts upon $\left( \omega ^{i}\right) $%
\begin{equation}
g\left( \omega ^{i}\right) =\left( \omega ^{i\prime }\right) h\left( \omega
^{i},g\right)   \label{2.5}
\end{equation}%
with $h\left( \omega ^{i},g\right) \in SO\left( 3\right) $.

To define a spin structure on $\mathcal{C}$ means to construct a principal
fiber bundle over $\mathcal{C}_{b}$ with typical fiber $Spin(3)$ (a double
cover of $\mathcal{C}$), such that the image of its transition functions 
\cite{spingeo} are the transition functions of $SO\left( 3\right) $. It is
known that a $SO\left( n\right) $ bundle over some base admits a spin-bundle
extension if and only if the second Stiefel-Whitney homology class, with $%
\mathbf{Z}_{2}$ coefficients, $w_{2}$ is trivial. However $SU(3)/SO(3)$ is
the Wu manifold which is known \cite{Wu} \cite{Atlas} not to be a $Spin$
manifold, nor a $Spin^{c}$ manifold. It is however a $Spin^{h}$ manifold 
\cite{spinh}. What this means is that to implement a spin structure in the $%
SO(3)$ principal bundle over $\mathcal{C}_{b}$, one should complement $%
\mathcal{C}$ by a Whitney sum with a second bundle $\mathcal{P}_{SU\left(
2\right) }^{\prime }$, over the same base $\mathcal{C}_{b}$, carrying a
quaternionic ($SU(2)$) structure, the fibers being the direct sum of the
fibers of the two bundles.

This has a simple interpretation in terms of the representations of the
transition functions. As stated before, the construction of the spin
structure implies the establishment of a matching of the transition
functions of the $SO(3)$ bundle with those of the corresponding spin bundle.
The transition functions are elements of the structure groups and therefore
the matching should occur at the level of its representations. The set $%
\left\{ R_{i}\right\} $ of $SO\left( 3\right) $ generators, as a subgroup of 
$SU\left( 3\right) $, carries a spin one representation, which does not
match the spin $1/2$ of the spin structure. Therefore one needs an
additional $SU\left( 2\right) $ spin $1/2$, which, coupled with the first
one, matches the spin one of the $SO\left( 3\right) $ bundle in the $%
SU\left( 3\right) $ triplet.

Use a creation operator representation of the Lie algebra of $SU\left(
3\right) $. Let $\left\{ a_{i}^{\dag },a_{i};i=1,2,3\right\} $ be a set of
bosonic creation and annihilation operators. Then the generators of the $%
SU\left( 3\right) $ algebra have the representation%
\begin{equation}
R_{i}=\epsilon _{ijk}a_{j}^{\dag }a_{k};\;U_{i}=i\left\vert \epsilon
_{ijk}\right\vert a_{j}^{\dag }a_{k};\;D_{1}=i\left( a_{1}^{\dag
}a_{1}-a_{2}^{\dag }a_{2}\right) ;\;D_{2}=\frac{i}{\sqrt{3}}\left(
a_{1}^{\dag }a_{1}+a_{2}^{\dag }a_{2}-2a_{3}^{\dag }a_{3}\right) ,
\label{2.6}
\end{equation}%
a $SU\left( 3\right) $ triplet being formed by $3$ spin-one states%
\begin{equation}
\left\vert 1\right) =\frac{1}{\sqrt{2}}\left( a_{1}^{\dag }+ia_{2}^{\dag
}\right) \left\vert 0\right\rangle ;\;\left\vert -1\right) =\frac{1}{\sqrt{2}%
}\left( a_{1}^{\dag }-ia_{2}^{\dag }\right) \left\vert 0\right\rangle
;\;\left\vert 0\right) =a_{3}^{\dag }\left\vert 0\right\rangle   \label{2.7}
\end{equation}%
Now one has to match these states by composing an observable spin $\frac{1}{2%
}$ state in the real Lorentzian fiber with another spin $\frac{1}{2}$ state
of the auxiliary $\mathcal{P}_{SU\left( 2\right) }^{\prime }$ bundle. Notice
however that the three states in (\ref{2.7}) not only transform under the $%
R_{i}$ rotation generators, but also have nontrivial transformation
properties under the other $SU\left( 3\right) $ generators, in particular $%
U_{1}$ and $U_{2}$. This must \ also be matched by the composite states.

Let $c_{i}^{\uparrow \dag }$ and $c_{j}^{\downarrow \dag }$ be the spin up
and spin down creator operators of the observable state in the real
Lorentzian fiber. And $b_{i}^{\uparrow \dag }$ and $b_{j}^{\downarrow \dag }$
the corresponding operators of the auxiliary $\mathcal{P}_{SU\left( 2\right)
}^{\prime }$ bundle. The $i$ and $j$ labels stand for the additional quantum
numbers needed to match the $SU\left( 3\right) $ transformation properties
of the states. The following scheme satisfies the matching requirements%
\begin{equation}
\left\vert 1\right) \rightarrow b_{i}^{\uparrow \dag }c_{i}^{\uparrow \dag
}\left\vert 0\right\rangle ;\;\left\vert -1\right) \rightarrow
b_{j}^{\downarrow \dag }c_{j}^{\downarrow \dag }\left\vert 0\right\rangle
;\;\left\vert 0\right) \rightarrow \left( b_{i}^{\uparrow \dag
}c_{j}^{\downarrow \dag }+b_{j}^{\downarrow \dag }c_{i}^{\uparrow \dag
}\right) \left\vert 0\right\rangle   \label{2.8}
\end{equation}%
Notice that the "internal" quantum numbers $i$ and $j$ of the spinor states $%
c_{i}^{\uparrow \dag }$ and $c_{j}^{\downarrow \dag }$ may be different for
different spin projections, showing the chiral nature of this quantum
number. The spinor doublets in the real Lorentzian fiber are thus endowed
with a doublet of "internal" quantum numbers. This might superficially look
like the emergence of an additional $SU\left( 2\right) $ structure. However
these quantum numbers are matching a structure acted upon by the $U_{i}$
operators which are not a closed $SU\left( 2\right) $ structure, but $%
SU\left( 3\right) $ operators. Therefore this emergent structure that will
tentatively be called "flavor" is not $SU\left( 2\right) $ but a doublet of
"flavor $SU\left( 3\right) $". This $SU\left( 3\right) $ structure, that at
the complex spacetime level is an exact symmetry, may connect states with
different $p^{2}=P^{\mu }P_{\mu }$values (the real mass square) because they
are eigenvalues of $P^{2}=P_{\mu }P^{\mu }+H_{\mu }H^{\mu }$, not of $p^{2}$
and from the commutation relations in (\ref{2.1}) one sees how the $U_{i}$
operators permute the $P_{\mu }$ and $H_{\mu }$ operators. Summarizing: an
exact complex symmetry may appear as a broken symmetry in the real
Lorentzian fibers. An additional feature of the (\ref{2.8}) structure is the
existence of a spin zero state%
\[
\left( b_{i}^{\uparrow \dag }c_{j}^{\downarrow \dag }-b_{j}^{\downarrow \dag
}c_{i}^{\uparrow \dag }\right) \left\vert 0\right\rangle 
\]

With the additional $\mathcal{P}_{SU\left( 2\right) }^{\prime }$ bundle,
spinor states are consistently constructed at each point of the coset space $%
SU(3)/SO(3)$. Therefore a second consequence is a degeneracy of the spinor
states parametrized by the points of the coset space. This is a $5-$%
dimensional space\footnote{%
Or $6-$dimensional if instead of $SU(3)/SO(3)$ one consider $U(3)/SO(3)$.}.
A detailed characterization of this "color" space and its coordinates (\ref%
{2.4}) is important, particularly if the $SU(3)$ symmetry, in its double
role, is gauged.

Summarizing: If the real Poincar\'{e} group is a symmetry of the real
Lorentzian fibers and $P_{\mathbb{C}}$ a symmetry of the full complex
spacetime, consistency of the representations generates a $U\left( 1\right)
\oplus SU\left( 3\right) $ structure with $SU\left( 3\right) $ acting in a
double role, both as generating additional doublet quantum numbers as a
broken $P_{\mu }P^{\mu }$ symmetry and generating a degeneracy of massive
spinor states parametrized by the base manifold.

\textbf{Acknowledgments}

Partially supported by Funda\c{c}\~{a}o para a Ci\^{e}ncia e a Tecnologia
(FCT), project UIDB/04561/2020: https://doi.org/10.54499/UIDB/04561/2020


\begin{thebibliography}{99}
\bibitem{Einstein} A. Einstein; \textit{A generalization of the relativistic
theory of gravitation}, Annals of Math. 46 (1945) 578-584.

\bibitem{Kunstatter} G. Kunstatter and R. Yates; \textit{The geometrical
structure of a complexified theory of gravitation}, J. Phys. A: Math Gen. 14
(1981) 847-854.

\bibitem{Plebanski} J. Plebanski; \textit{Some solutions of complex Einstein
equations}, J. Math. Phys. 16 (1975) 2395-2402.

\bibitem{Penrose-2} R. Penrose; \textit{The twistor approach to space-time
structures}, in \textit{100 Years of Relativity}, A. Ashtekar (Ed.) pp.
465-505, World Scientific, Singapore 2005.

\bibitem{Esposito-1} G. Esposito; \textit{From spinor geometry to complex
general relativity, }Int. J. of Geometric Methods in Modern Physics 2 (2005)
675-731.

\bibitem{Brown} E. H. Brown; \textit{On the complex structure of the universe%
}, J. Math. Phys. 7 (1966) 417-425.

\bibitem{Newman} E. T. Newman; \textit{Maxwell's equations and complex
Minkowski spac}e; J. Math. Phys. 14 (1973) 102-103.

\bibitem{Esposito-2} G. Esposito; C\textit{omplex geometry of Nature and
general relativity}, Kluwer Acad. Press, Dordrecht 2002.

\bibitem{IJMPA} R. Vilela Mendes; \textit{Space times over normed division
algebras, revisited}, Int. J. Modern Physics A 35 (2020) 2050055.

\bibitem{Barut} A. O. Barut; \textit{Complex Lorentz group with a real
metric: Group structure}, J. Math. Phys. 5 (1964) 1652-1656.

\bibitem{Ungar} A. A. Ungar; \textit{The abstract complex Lorentz
transformation group with real metric. II. The invariance group of the form} 
$\Vert t\Vert ^{2}-\parallel x\parallel ^{2}$, J. Math. Phys. 35 (1994)
1881-1913.

\bibitem{Kursunoglu} B. Kursunoglu; \textit{New symmetry group for
elementary particles. I. Generalization of Lorentz group via electrodynamics}%
, Phys. Rev. 135 (1964) B761-B768.

\bibitem{dark} R. Vilela Mendes; \textit{T-violation and the dark sector},
Modern Physics Letters A 38 (2023) 2350165.

\bibitem{spingeo} H. B. Lawson and M-L. Michelsohn; \textit{Spin geometry},
Princeton Univ. Press, Princeton 1989.

\bibitem{Wu} W. T. Wu; \textit{Classes caract\'{e}ristiques et i-carr\'{e}s
d'une vari\'{e}t\'{e}}, C. R. Acad. Sci., Paris, 230 (1950), 508-509

\bibitem{Atlas} D. Crowley; \textit{5-manifolds: 1-connected}, Bulletin of
the Manifold Atlas (2011) 49-55.

\bibitem{spinh} M. Albanese and A. Milivojevi\'{c}; \textit{Spin}$^{h}$%
\textit{\ and further generalisations of spin}, Journal of Geometry and
Physics 164 (2021) 104174, 184 (2023) 104709.
\end{thebibliography}
\end{document}